\documentclass[final,3p,times]{elsarticle}

\usepackage{amssymb}
\usepackage{amsmath}

\newcommand{\pp}[1]{$p$+$p$}

\journal{Physics Letters B}

\begin{document}

\begin{frontmatter}

\title{Measurements of charged-particle multiplicity dependence of higher-order net-proton cumulants in $p$+$p$ collisions at $\sqrt{s}~=$~200~GeV from STAR at RHIC}

\author{
M.~I.~Abdulhamid$^{4}$,
B.~E.~Aboona$^{56}$,
J.~Adam$^{16}$,
L.~Adamczyk$^{2}$,
J.~R.~Adams$^{41}$,
I.~Aggarwal$^{43}$,
M.~M.~Aggarwal$^{43}$,
Z.~Ahammed$^{63}$,
E.~C.~Aschenauer$^{6}$,
S.~Aslam$^{28}$
% ...
}

\address{\rm{(STAR Collaboration)}}

% ...
\address{$^{2}$AGH University of Krakow, FPACS, Cracow 30-059, Poland}
% ...
\address{$^{4}$American University in Cairo, New Cairo 11835, Egypt}
% ...
\address{$^{6}$Brookhaven National Laboratory, Upton, New York 11973}
% ...
\address{$^{16}$Czech Technical University in Prague, FNSPE, Prague 115 19, Czech Republic}
% ...
\address{$^{28}$Indian Institute Technology, Patna, Bihar 801106, India}
% ...
\address{$^{41}$The Ohio State University, Columbus, Ohio 43210}
% ...
\address{$^{43}$Panjab University, Chandigarh 160014, India}
% ...
\address{$^{56}$Texas A\&M University, College Station, Texas 77843}
% ...
\address{$^{63}$Variable Energy Cyclotron Centre, Kolkata 700064, India}
% ...

\begin{abstract}
We report on the charged-particle multiplicity dependence of net-proton cumulant ratios up to sixth order from $\sqrt{s}$~=~200~GeV $p$+$p$ collisions at the Relativistic Heavy Ion Collider (RHIC). The measured ratios $C_{4}/C_{2}$, $C_{5}/C_{1}$, and $C_{6}/C_{2}$ decrease with increased charged-particle multiplicity and rapidity acceptance. Neither the Skellam baselines nor PYTHIA8 calculations account for the observed multiplicity dependence. In addition, the ratios $C_{5}/C_{1}$ and $C_{6}/C_{2}$ approach negative values in the highest-multiplicity events, which implies that thermalized QCD matter may be formed in $p$+$p$ collisions. 
\end{abstract}


\end{frontmatter}

\section{Introduction}
Exploring the quantum chromo-dynamics (QCD) phase structure is one of the ultimate goals in heavy-ion collision experiments.
The QCD phase diagram is characterized in terms of baryon chemical potential ($\mu_{\rm B}$) for the x-axis and temperature ($T$) for the y-axis. 
In the conjectured structure of the QCD phase diagram, 
there is a quark-gluon plasma (QGP) phase
in the high temperature region~\cite{Niida:2021wut} where the thermalized QCD matter is formed, 
while a hadron gas phase exists at lower temperature and higher baryon density~\cite{Bzdak:2019pkr}. 
However, the transition between the two phases is not well-understood.
According to lattice QCD calculations, the phase transition at $\mu_{\rm B}/T<2$ is likely a smooth crossover~\cite{Borsanyi:2018grb,Aoki:2006we}. 
Model calculations predict a first-order phase transition in the finite $\mu_{\rm B}$ region~\cite{Bowman:2008kc} as well as a QCD critical point~\cite{StephanovMA:2011}.

Cumulants of conserved charges are believed to be sensitive to the QCD phase structure~\cite{correlation}.
Various orders of cumulants ($C_{n},\, n\leq6$), of net-proton, net-kaon, and net-charge multiplicity 
distributions have been measured by the ALICE, STAR, HADES, and NA61/SHINE collaborations~\cite{net_proton,net_charge,STAR:2017tfy,STAR:2021iop,STAR:2020tga,Mackowiak-Pawlowska:2021pub,Adamczewski-Musch:2020slf,ALICE:2019nbs}.
The net-proton $C_{4}/C_{2}$ in the Beam Energy Scan program phase I (BES-I) ($7.7\leq\sqrt{s_{\rm NN}}\leq200$~GeV) 
at RHIC indicates a non-monotonic collision-energy dependence with $3.1\sigma$ significance, 
which hints at the existence of a critical point at $\sqrt{s_{\rm NN}}<20$~GeV ~\cite{Stephanov:2011pb}. 
The $C_4/C_2$ value from the fixed-target experiment for 3~GeV collisions is consistent with hadron transport model calculations~\cite{STAR:2021fge}. 
For precise measurements at $3<\sqrt{s_{\rm NN}}<20$~GeV, the Beam Energy Scan program phase II (BES-II) was carried out from 2019 to 2021 both in the collider mode ($7.7\leq\sqrt{s_{\rm NN}}\leq19.6$~GeV) and the fixed-target mode ($3.2\leq\sqrt{s_{\rm NN}}\leq7.7$~GeV). 

Higher-order cumulant ratios have been measured by STAR for Au+Au collisions at $\sqrt{s_{\rm NN}}=$~200~GeV~\cite{STAR:2021rls}. In particular, the ratios $C_{6}/C_{2}$ were found to be systematically negative from peripheral to central collisions. Albeit with large uncertainties, the negative values are qualitatively consistent with QCD models and lattice QCD calculations~\cite{Friman,Fu:2021oaw,BazavovmuT:2020} which seems to favor a smooth crossover transition.

This Letter reports the net-proton cumulant ratios, $C_{2}/C_{1}$, $C_{3}/C_{2}$, $C_{4}/C_{2}$, $C_{5}/C_{1}$, 
and $C_{6}/C_{2}$ measured in $p$+$p$ collisions at $\sqrt{s}=$~200~GeV.
Experimental results from RHIC and the Large Hadron Collider~(LHC) suggest that the QGP could be formed in high-multiplicity $p$+$p$ collisions~\cite{CMS:2010ifv,ATLAS:2015hzw,CMS:2016fnw,ALICE:2016fzo}.
This scenario can be tested by measuring the multiplicity dependence of higher-order cumulant ratios.
Although collectivity~\cite{CMS:2010ifv,ATLAS:2015hzw}, collective motion of final particles with different masses such as the anisotropy flow $v_{2}$, and strangeness enhancement~\cite{ALICE:2016fzo}, due to abundant production of gluons, have been observed in $p$+$p$ collisions, 
it does not necessarily indicate that the thermalized QCD drops of matter have been created. On the other hand, higher-order cumulants of net-proton distributions offer more sensitive probe for the nature of thermalization in such collisions. This will be done by comparing various orders of cumulants between data and calculations from both thermal model and transport model~\cite{Gupta:2022phu}.

\section{Dataset}
\begin{figure}[htbp]
\begin{center}
\hspace*{-0.8cm}
\includegraphics[width=90mm]{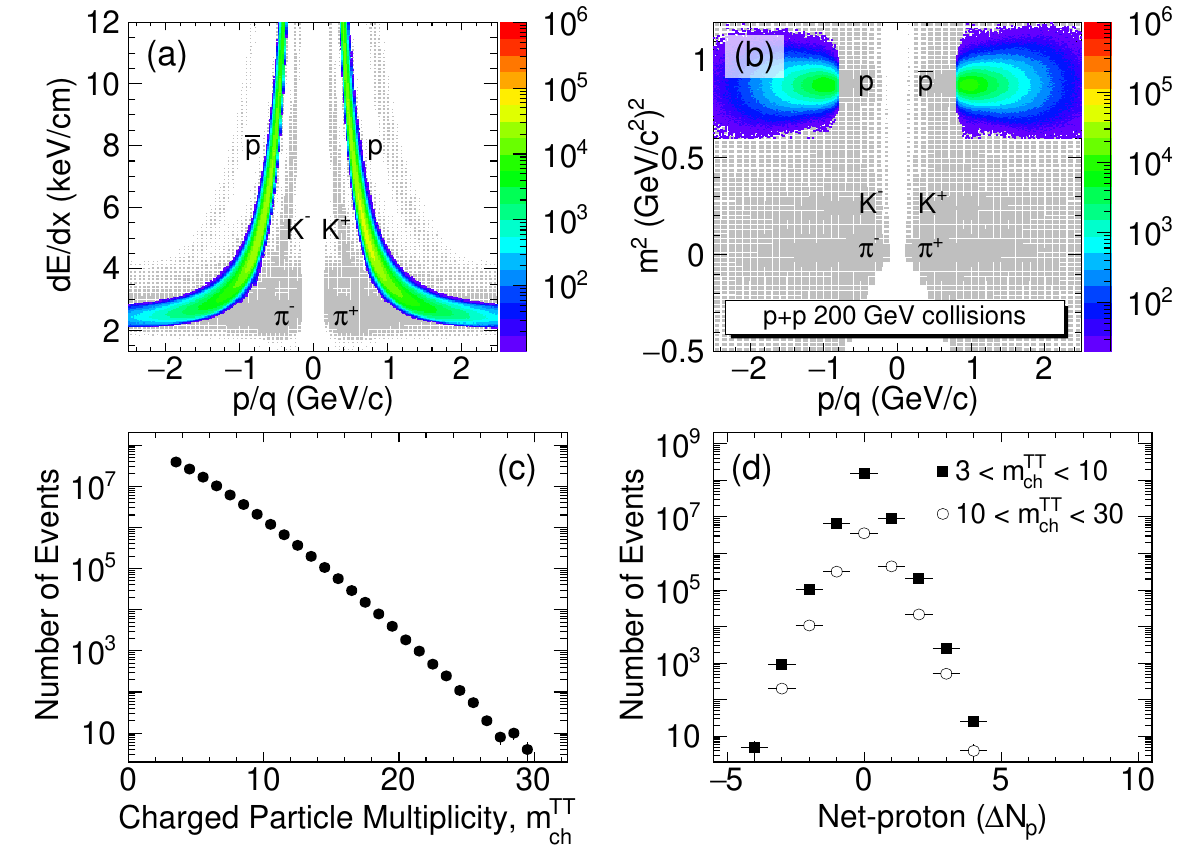}
\end{center}
\vspace{-0.7cm}
\caption{
   (a) Correlations between the energy loss, d$E/$d$x$, of charged tracks measured by the TPC and momentum divided by electric charge.  
   (b) Correlations between the mass squared, $m^{2}$, measured by the TOF and the ratio of momentum to the electric charge, $p/q$. 
   Colored bands represent the identified protons and antiprotons used in the analysis.
   (c) Charged-particle multiplicity distribution.
   (d) Event-by-event net-proton multiplicity distributions for $|y|<0.5$ and $0.4<p_{\rm{T}}<2.0$~GeV/$c$ 
   at two ranges of charged particle multiplicity as indicated in the legend.  
}
\label{fig:Fig1}
\end{figure}

The data set was taken in 2012 using a minimum-bias trigger, with 220 million events analyzed.
All data were measured using the Time Projection Chamber (TPC) and Time of Flight (TOF) detector at the STAR experiment~\cite{KHAckermann:2003}.
The collision vertex is required to be within $30$~cm of the detector center and within $2$~cm in the transverse direction relative to the beamline.
Pileup events are suppressed by requiring the difference in the vertex position along the beamline measured by
the TPC and the Vertex Position Detector to be within $\pm3$~cm, and by requiring that tracks are matched to the TOF hits.
Protons and antiprotons are analyzed in the transverse momentum range $0.4<p_{\rm{T}}<2.0$~GeV/$c$ and in the midrapidity acceptance $|y|<0.5$.
Figure~\ref{fig:Fig1} (a) shows the ionization energy loss (d$E/$d$x$) as a function of the momentum of charged particles, as measured by the TPC.
Figure~\ref{fig:Fig1} (b) depicts correlations between mass squared ($m^{2}$) measured by the TOF and the momentum measured by the TPC.
Protons and antiprotons are identified by using both d$E/$d$x$ and $m^{2}$. 
The distance of closest approach of a reconstructed track to the collision vertex is required to be less than $1$~cm to suppress the contribution from secondary protons from weak decays.
Cumulants are calculated at each bin of measured charged-particle multiplicity, $m_{\rm ch}^{\rm TT}$, 
which is defined as charged tracks having hits both in the TPC and TOF at midrapidity. 
Protons and antiprotons are excluded from $m_{\rm ch}^{\rm TT}$ in order to suppress self-correlations~\cite{Autocorrelations:2013xluo}. 
The $m_{\rm ch}^{\rm TT}$ distribution is shown in Fig.~\ref{fig:Fig1} (c). The event-by-event net-proton number distributions for two ranges of $m_{\rm ch}^{\rm TT}$ are shown in Fig.~\ref{fig:Fig1} (d). The TOF hit requirement is designed to suppress the contribution from collision pileup events in high-luminosity \pp\, collisions. Since the efficiency of the TOF is well understood in the experiment~\cite{STAR:2021iop}, $m_{\rm ch}^{\rm TT}$ is converted to the corresponding multiplicity in the TPC, $m_{\rm ch}^{\rm TPC}$, in subsequent discussions, as the new results will be finally compared with previous measurements from Au+Au collisions~\cite{MSAbdallah:2021May}.

\section{Observables}
The $n$th-order cumulant is defined by the $n$th derivatives of the cumulant-generating function. 
Explicitly, cumulants are expressed in Eqs.~(\ref{cneqs1})-(\ref{cneqs6}).
\begin{eqnarray}
\label{cneqs1}
C_{1}&=&\langle N\rangle, \\
\label{cneqs2}
C_{2}&=&\langle(\delta N)^{2}\rangle, \\
\label{cneqs3}
C_{3}&=&\langle(\delta N)^{3}\rangle, \\
\label{cneqs4}
C_{4}&=&\langle(\delta N)^{4}-3\langle(\delta N)^{2}\rangle \rangle^{2}, \\
\label{cneqs5}
C_{5}&=&\langle(\delta N)^{5}\rangle-10\langle(\delta N)^{3}\rangle \langle(\delta N)^{2}\rangle, \\ 
C_{6}&=&\langle(\delta N)^{6}\rangle-15\langle(\delta N)^{4}\rangle \langle(\delta N)^{2}\rangle \nonumber \\
\label{cneqs6}
&-&10\langle(\delta N)^{3}\rangle^{2} +30\langle(\delta N)^{2}\rangle^{3}, 
\end{eqnarray}
where $\langle\delta N\rangle=N-\langle N\rangle$, $N$ is the number of particles in one event, and the angle brackets indicate the average over all events.

The cumulant ratios, $C_{2}/C_{1}$, $C_{3}/C_{2}$, $C_{4}/C_{2}$, $C_{5}/C_{1}$, and $C_{6}/C_{2}$, are employed to cancel the trivial volume dependence of cumulants~\cite{CumulantEqWithVolume:2009mch}. 
Neutrons cannot be measured in the STAR experiment, hence net-proton number is measured as a proxy for net-baryon number~\cite{eff_kitazawa}. 
A Skellam distribution, which is defined as the difference between two independent Poisson distributions, is employed as a statistical baseline. 
The odd- and even-order cumulants of the Skellam distribution are expressed by the difference and the sum of the averaged values of protons and antiprotons, respectively. 
Consequently, the Skellam baselines for $C_{4}/C_{2}$, $C_{5}/C_{1}$, and $C_{6}/C_{2}$ are always unity, 
while the corresponding baselines for $C_{2}/C_{1}$ and $C_{3}/C_{2}$ depend on the averaged value of protons and antiprotons.

\section{Analysis methods}
In order to account for detector efficiency effects present in cumulants calculated from Fig.~\ref{fig:Fig1} (d), efficiency corrections are applied to the measured cumulants including luminosity, acceptance, and charged-particle multiplicity dependencies~\cite{binomial_breaking}. 
The detector efficiencies are assumed to follow binomial distributions~\cite{eff_kitazawa,eff_koch,tsukuba_eff_separate,eff_psd_volker,eff_xiaofeng,Luo:2018ofd,eff_psd_kitazawa,Nonaka:2017kko,Nonaka:2023sxt}.
Statistical uncertainties are estimated by the bootstrap method~\cite{Luo:2017faz,Luo:2018ofd}.
Systematic uncertainties are estimated by varying the selection ranges for track quality, particle identification, luminosity, and by varying the detector efficiencies for the efficiency corrections. 
The luminosity ranges from 2 kHz to 15 kHz for the coincidence rates for the Zero Degree Calorimeters, which is divided into 10 groups to estimate the variations of the cumulants. 
A Barlow test was performed in order to remove contribution from statistical uncertainties~\cite{Barlow:2002yb}. 
The systematic uncertainties from each source are 0.29\%, 0.32\%, 0.38\%, and 0.69\%, respectively, for multiplicity-averaged results of $C_{4}/C_{2}$.
The total systematic uncertainties are 0.90\%, 7.8\%, and 8.4\% for $C_{4}/C_{2}$, $C_{5}/C_{1}$, and $C_{6}/C_{2}$, respectively. 

\section{Results and discussions}
\begin{figure}[htbp]
\begin{center}
\includegraphics[width=120mm]{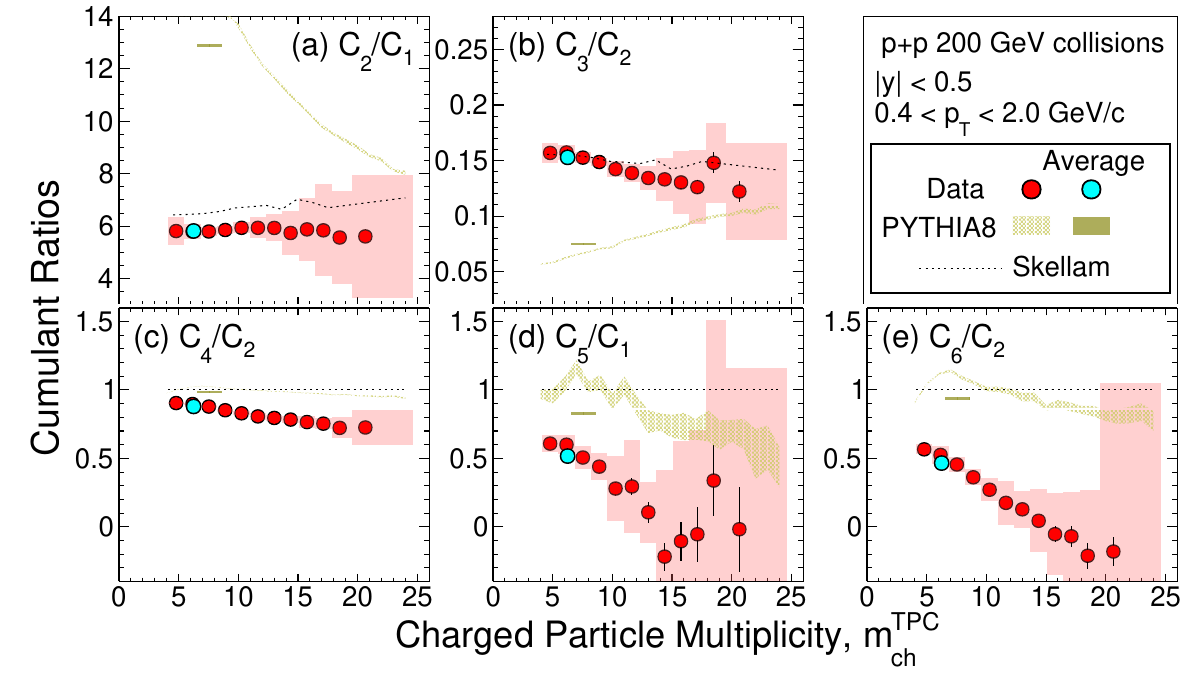}
\end{center}
\vspace{-0.7cm}
\caption{
    Net-proton cumulant ratios, (a) $C_{2}/C_{1}$, (b) $C_{3}/C_{2}$, (c) $C_{4}/C_{2}$, (d) $C_{5}/C_{1}$, and (e) $C_{6}/C_{2}$ as a function of charged-particle multiplicity from $\sqrt{s}~=$~200~GeV $p$+$p$ collisions. Black solid lines and red bands represent the statistical and systematic uncertainties, respectively. 
    Cyan points represent event averages for $ 3<m_{\rm ch} < 30$, and they are plotted at the corresponding value of $m_{\rm ch}^{\rm TPC}$. The uncertainties on the cyan points are smaller than the marker size.
    The Skellam baselines are shown as dashed lines.
    The results of the PYTHIA8 calculations are shown by hatched-golden bands. The golden bands at $m_{\rm ch}^{\rm TPC}\approx 6$ are the results from the PYTHIA8 calculations averaged over multiplicities. 
}
\label{fig:FigforPaper_Ratio}
\end{figure}

Charged-particle multiplicity dependence for the net-proton cumulant ratios,  $C_{4}/C_{2}$, $C_{5}/C_{1}$, and $C_{6}/C_{2}$ are shown in Fig.~\ref{fig:FigforPaper_Ratio} as red dots. Vertical bars represent statistical uncertainties while the red bands represent systematic uncertainties.
The main source of the large systematic uncertainties for each charged-particle multiplicity bin is the luminosity. It constitutes 50\% out of total systematic uncertainty for $C_{4}/C_{2}$ at $m_{\rm ch}^{\rm TPC}=4$, which increases up to above 90\% at $m_{\rm ch}^{\rm TPC}>18$. The effect is more prominent for $C_{5}/C_{1}$ and $C_{6}/C_{2}$, greater than 90\% in most $m_{\rm ch}^{\rm TPC}$ bins.
Calculations from the Skellam baseline and PYTHIA8 model~\cite{Sjostrand:2014zea} are shown as black dashed lines and purple bands, respectively. Approximately 800 million PYTHIA8 events, with the option of SoftQCD, are generated for the comparison. Figure \ref{fig:FigforPaper_Ratio} indicates all ratios are decreasing with increasing multiplicity, except for  $C_2/C_1$. Although the PYTHIA8 model shows multiplicity dependence, it does not provide quantitative description of data. 
For the higher-order ratios $C_{4}/C_{2}$, $C_{5}/C_{1}$, and $C_{6}/C_{2}$, the model (purple squares) also overpredicts the multiplicity-averaged data values (cyan solid circles). 
As reported in Ref.~\cite{STAR:2022vlo}, thermodynamic model calculations from both lattice QCD~\cite{Borsanyi:2018grb,BazavovmuT:2020} and functional renormalization group (FRG)~\cite{Fu:2021oaw} predict a special ordering of the higher-order baryon-number susceptibility ratios: $\chi^{\rm B}_4/\chi^{\rm B}_2>\chi^{\rm B}_5/\chi^{\rm B}_1>\chi^{\rm B}_6/\chi^{\rm B}_2$. The multiplicity-averaged ratios in Fig.~\ref{fig:FigforPaper_Ratio} show the hierarchy $C_4/C_2>C_5/C_1>C_6/C_2$, where the significance of the first and second inequalities is 8.7~$\sigma$ and 0.9~$\sigma$, respectively. At high charged-particle multiplicity bins $m_{\rm ch}^{\rm TPC}>12$, both values of $C_5/C_1$ and $C_6/C_2$ are consistent with zero within uncertainties, which hints at a possible sign change in these ratios at even higher multiplicity. These observations are consistent with the thermodynamic model expectation for the formation of thermalized QCD matter in high-multiplicity $p$+$p$ collisions. We discuss this point later when comparing with results from 200~GeV Au+Au collisions.
Note that the PYTHIA8 calculations fail to reproduce the hierarchy in the ratios and no hint of a sign change is observed. At LHC energies, due to multi-parton interactions in \pp\, collisions, the color-reconnection (CR) mechanism is used to mimic collective excitation for strangeness and heavy-quark production~\cite{Gross:2022hyw,ALICE:2016fzo,ALICE:2012pet,ALICE:2022wpn,CMS:2020qul,ATLAS:2015hzw}. 
We confirmed that PYTHIA8 calculations with CR included do not have any significant effect on the net-proton higher-order cumulant ratios from the 200~GeV \pp\, collisions.

\begin{figure}[htbp]
\begin{center}
\includegraphics[width=90mm]{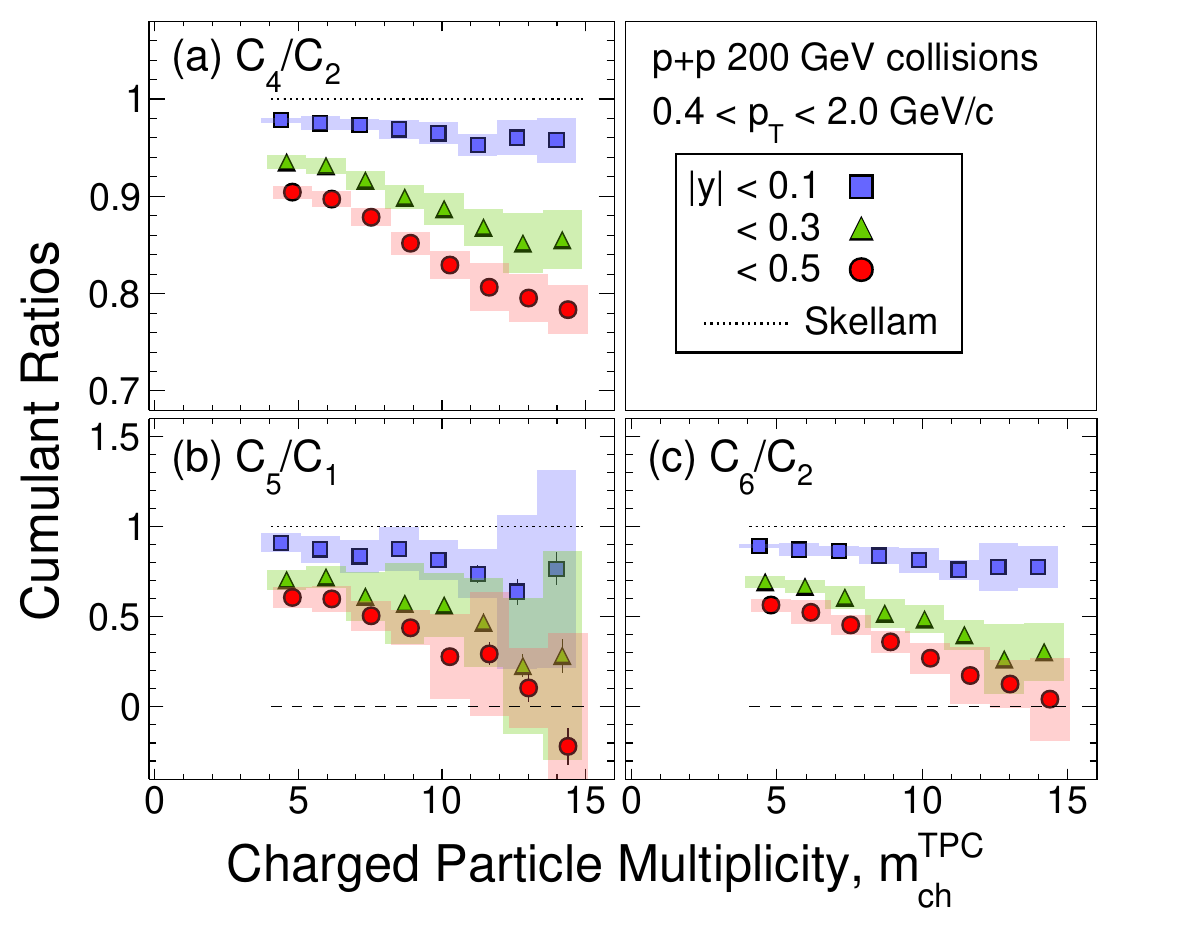}
\end{center}
\vspace{-0.7cm}
\caption{Rapidity-acceptance dependence of the net-proton cumulant ratios, (a) $C_{4}/C_{2}$, (b) $C_{5}/C_{1}$, and (c) $C_{6}/C_{2}$ shown 
as a function of the charged-particle multiplicity, $m_{\rm ch}^{\rm TPC}$, from $\sqrt{s}=$ 200~GeV $p$+$p$ collisions. 
Solid squares, triangles, and circles represent the results with $|y|<0.1$, $0.3$, and $0.5$, respectively. 
Black solid lines and colored bands show the statistical and systematic uncertainties.
The data points and their uncertainties for $|y|<0.1$ and $0.3$ are slightly shifted for clarity.
The Skellam baselines are shown by dashed lines.}
\label{fig:ydep}
\end{figure}

Studying the rapidity dependence of the cumulant ratios may shed light on the time evolution of the collision dynamics~\cite{Sakaida:2017rtj}. The rapidity-window dependence of the net-proton cumulant ratios vs. multiplicity is shown in Fig.~\ref{fig:ydep}, with the colored markers representing data from three rapidity bins. In order to avoid overlapping uncertainties, data are plotted up to $m_{\rm ch}^{\rm TPC}=$ 14. 
The higher-order ratios $C_{4}/C_{2}$, $C_{5}/C_{1}$, and $C_{6}/C_{2}$ show that the larger the rapidity acceptance, the more the ratios deviate from the corresponding ratio of the Skellam distribution (dashed lines) and the values are largely decreasing with increasing multiplicity. Both features imply the strongest correlation in the largest rapidity window. A similar rapidity-window dependence has also been observed in Au+Au collisions at RHIC~\cite{STAR:2021rls}. In addition, in the widest rapidity bin, $|y|<0.5$, the ratios $C_{5}/C_{1}$ and $C_{6}/C_{2}$ approach negative values at the highest multiplicity, indicating that thermalized QCD matter may be created in very high-multiplicity events in \pp\, collisions at RHIC, as we noted for Fig.~\ref{fig:FigforPaper_Ratio}. None of the observations are reproduced by the QCD-inspired event generator PYTHIA8, which implies a lack of collective excitation dynamics in the model.
Please note that the baryon number conservation could also lead to the decrease of higher-order cumulant ratios, especially for $C_{5}/C_{1}$ and $C_{6}/C_{2}$, and the effect depends on the detector acceptance~\cite{Bzdak:2012an}. Using the PYTHIA8 model and STAR acceptance, we find that both the ratios decrease as charged multiplicity increases, but the rate of the decrease is slower than that of data. More importantly, these ratios remain to be positive even for the highest multiplicity events. (More details can be found in the supplemental materials.) 
We also note that the baryon number conservation is present in PYTHIA8. Nevertheless, the ratios from PYTHIA8 does not show negative signs of $C_{5}/C_{1}$ and $C_{6}/C_{2}$ and the decreasing trend with respect to the multiplicity is not as prominent as the experimental results.

\begin{figure*}[htbp]
\begin{center}
\includegraphics[width=160mm]{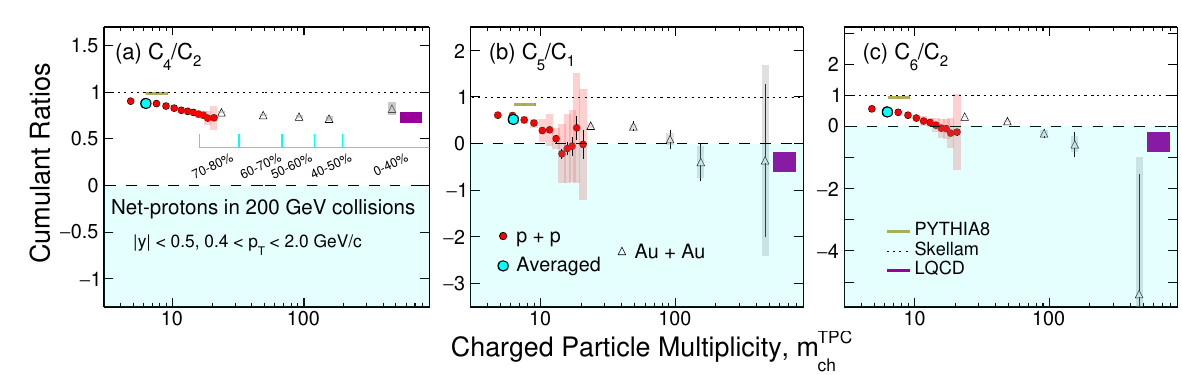}
\end{center}
\vspace{-0.7cm}
\caption{
Net-proton cumulant ratios, (a) $C_{4}/C_{2}$, (b) $C_{5}/C_{1}$, and (c) $C_{6}/C_{2}$ as a function of charged-particle multiplicity for $p$+$p$ collisions and Au+Au collisions at 200~GeV for $|y|<0.5$ and $0.4<p_{\rm T}<2.0$~GeV/$c$. Cyan markers represent event averages for $3<m_{\rm ch}^{\rm TPC} <30$ from the \pp\, collisions.  
Results from Au+Au collisions are shown as triangles for the 0-40\%, 40-50\%, 50-60\%, 60-70\%, and 70-80\% centrality bins. 
Red and grey bands show the systematic uncertainties for $p$+$p$ collisions and Au+Au collisions, respectively. 
The $m_{\rm ch}^{\rm TPC}$ values are not corrected for the reconstruction efficiency of the TPC. 
The golden bands at $m_{\rm ch}^{\rm TPC}\approx 6$ represent the results from PYTHIA8 calculations averaged over multiplicity.
The Skellam baselines are shown in dotted lines.
The purple bands show corresponding susceptibility ratios of baryon number from lattice QCD calculations~\cite{BazavovmuT:2020}, where the multiplicity range is chosen arbitrary. 
}
\label{fig:plot1_ppnetp_March2021}
\end{figure*}

Net-proton ratios $C_{4}/C_{2}$, $C_{5}/C_{1}$, and $C_{6}/C_{2}$
are compared with those from Au+Au collisions at $\sqrt{s_{\rm NN}}=$~200~GeV~\cite{STAR:2021iop,STAR:2022vlo} in Fig.~\ref{fig:plot1_ppnetp_March2021}, shown as a function of the charged-particle multiplicity. The kinematic acceptance $0.4<p_{\rm T}<2.0$~GeV/$c$ and $|y|<0.5$ is used in both \pp\, and Au+Au collisions. The five data points for Au+Au collisions correspond to 0-40\%, 40-50\%, 50-60\%, 60-70\%, and 70-80\% centrality bins. 
The multiplicity averaged ratios from the \pp\, collisions seem to follow the trend in the peripheral Au+Au collisions. However, the slope of the multiplicity dependence in the \pp\, range, $5<m_{\rm ch}^{\rm TPC}<20$, is steeper than that in the range of $25<m_{\rm ch}^{\rm TPC}<460$, for Au+Au collisions. The steeper slope indicates that thermalized QCD matter are created more efficiently in the high multiplicity \pp\, collisions than that in heavy-ion collisions.
In particular, both the ratios $C_{5}/C_{1}$ and $C_{6}/C_{2}$ from \pp\, collisions decrease with multiplicity, approach negative values for the highest-multiplicity events. 
The ratios from Au+Au collisions at similar multiplicity region are positive, which could be attributed as the initial volume fluctuations in heavy-ion collisions~\cite{Skokov:2012ds,Braun-Munzinger:2016yjz,Sugiura:2019toh}.
The negative ratios in the context of lattice QCD calculations imply that even in the small \pp\, system, thermalized QCD matter may be created in the highest-multiplicity collisions, 
if the trend in multiplicity dependence of the cumulant ratios reflects the thermalization of the bulk matter as that in Au+Au collisions.
Similar phenomena have been observed in the multiplicity dependence of strange hadron and J/$\psi$ production~\cite{ALICE:2016fzo,ALICE:2012pet}, 
and long-range collective motion~\cite{CMS:2010ifv,ATLAS:2015hzw} in $p$+$p$ collisions at LHC energies. 
Overall, the high-order net-proton cumulant ratios from both 200~GeV $p$+$p$ and Au+Au collisions show a clear decreasing trend from low to high charged-particle multiplicity, and eventually reach values
consistent with lattice QCD calculations assuming $\mu_{\rm B}=$~25~MeV and $T=$~155~MeV~\cite{BazavovmuT:2020}, within uncertainties.
Measurements of the cumulant ratios at high center-of-mass energies or in larger collision system than $p$+$p$, (e.g. $p$+Au, $d$+Au, Zr+Zr, and Ru+Ru collision systems) could provide systematic information on the dynamics of the observed multiplicity dependence. 

\section{Summary}
In summary, we report the first measurements of higher-order cumulant ratios $C_{2}/C_{1}$, $C_{3}/C_{2}$, $C_{4}/C_{2}$, $C_{5}/C_{1}$ and $C_{6}/C_{2}$ of net-proton multiplicity distributions in $p$+$p$ collisions at $\sqrt{s}=$~200~GeV, as measured by the STAR detector at RHIC.
Both charged-particle multiplicity and rapidity-cut dependencies are reported.
It is found that the ratios are all below Skellam expectations. 
Overall, the net-proton cumulant ratios $C_{4}/C_{2}$, $C_{5}/C_{1}$ and $C_{6}/C_{2}$ decrease progressively between the 200~GeV $p$+$p$ collisions and Au+Au central collisions. 
Calculations from PYTHIA8~\cite{Sjostrand:2014zea} fail to reproduce the multiplicity dependence and the hierarchy observed in net-proton ratios $C_{4}/C_{2}$, $C_{5}/C_{1}$ and $C_{6}/C_{2}$.
The $C_5/C_1$ and $C_6/C_2$ values are consistent with zero within uncertainties at the highest multiplicity bins with the maximized rapidity acceptance. 
Negative values of $C_{5}/C_{1}$ and $C_{6}/C_{2}$ are predicted, according to lattice QCD calculations, for thermalized nuclear matter~\cite{Borsanyi:2018grb,Aoki:2006we}. 
These new results imply that the thermalized QCD drops of matter could be created in truely high multiplicity events of the $p$+$p$ collisions. 
Systematic measurements of the cumulant ratios from higher collision energies, in larger colliding systems, and wider acceptance in rapidity will provide important information on the underlying dynamics of high-order net-proton cumulants and the process of thermalization in high-energy collisions. 

\section*{Acknowledgements}
We thank the RHIC Operations Group and RCF at BNL, the NERSC Center at LBNL, and the Open Science Grid consortium for providing resources and support.  This work was supported in part by the Office of Nuclear Physics within the U.S. DOE Office of Science, the U.S. National Science Foundation, National Natural Science Foundation of China, Chinese Academy of Science, the Ministry of Science and Technology of China and the Chinese Ministry of Education, the Higher Education Sprout Project by Ministry of Education at NCKU, the National Research Foundation of Korea, Czech Science Foundation and Ministry of Education, Youth and Sports of the Czech Republic, Hungarian National Research, Development and Innovation Office, New National Excellency Programme of the Hungarian Ministry of Human Capacities, Department of Atomic Energy and Department of Science and Technology of the Government of India, the National Science Centre and WUT ID-UB of Poland, the Ministry of Science, Education and Sports of the Republic of Croatia, German Bundesministerium f\"ur Bildung, Wissenschaft, Forschung and Technologie (BMBF), Helmholtz Association, Ministry of Education, Culture, Sports, Science, and Technology (MEXT), Japan Society for the Promotion of Science (JSPS) and Agencia Nacional de Investigaci\'on y Desarrollo (ANID) of Chile.
\bibliographystyle{elsarticle-num} 
\bibliography{main}

\section*{Appendix: Supplementary material}
\subsection*{Effect of baryon number conservation}
As discussed in Ref.~\cite{Bzdak:2012an}, it is known that the baryon number conservation artificially decreases the higher-order cumulants of net-particle distribution. To study the effect on our measurements, we performed numerical calculations with the procedures as follows.

\begin{figure}[htbp]
\begin{center}
\includegraphics[width=140mm]{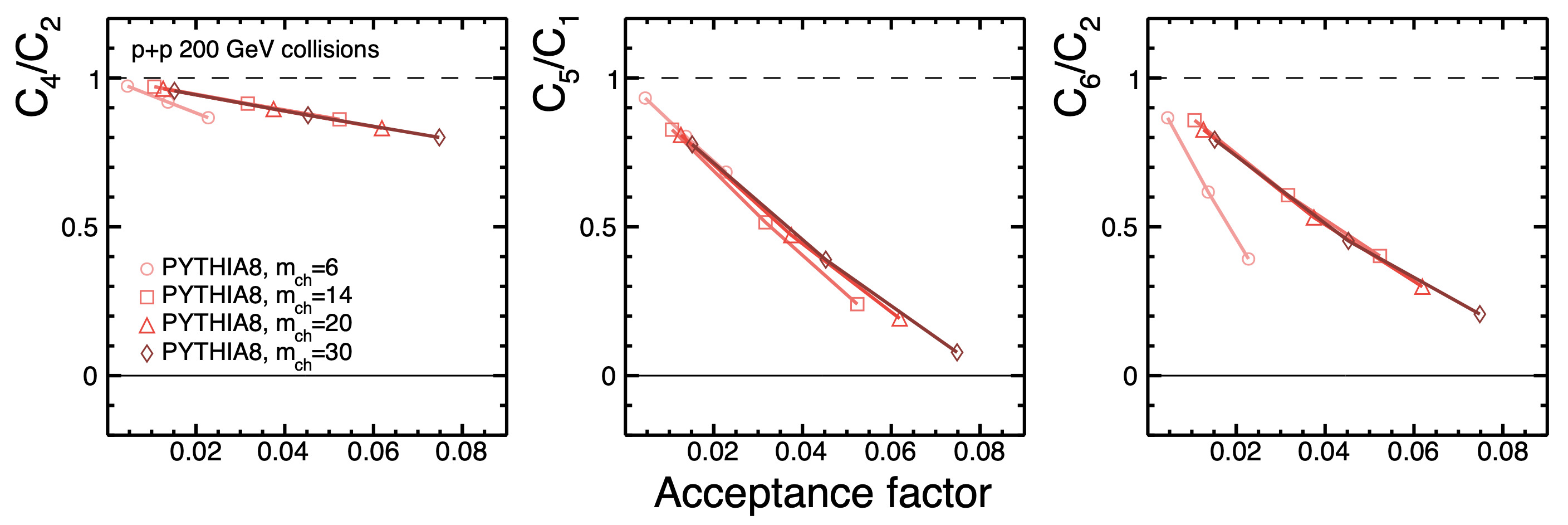}
\vspace{-0.5cm}
\end{center}
\caption{
Results from numerical calculations on net-proton $C_{4}/C_{2}$, $C_{5}/C_{1}$, and $C_{6}/C_{2}$ as a function of acceptance factor. Different markers represent four charged particle multiplicity bins chosen in PYTHIA8 calculations. Three data points for each line corresponds to the proton rapidity acceptance of $|y|<0.1$, $0.3$, and $0.5$.
}
\label{fig:Fig1_sup}
\end{figure}

\begin{figure}[htbp]
\begin{center}
\includegraphics[width=160mm]{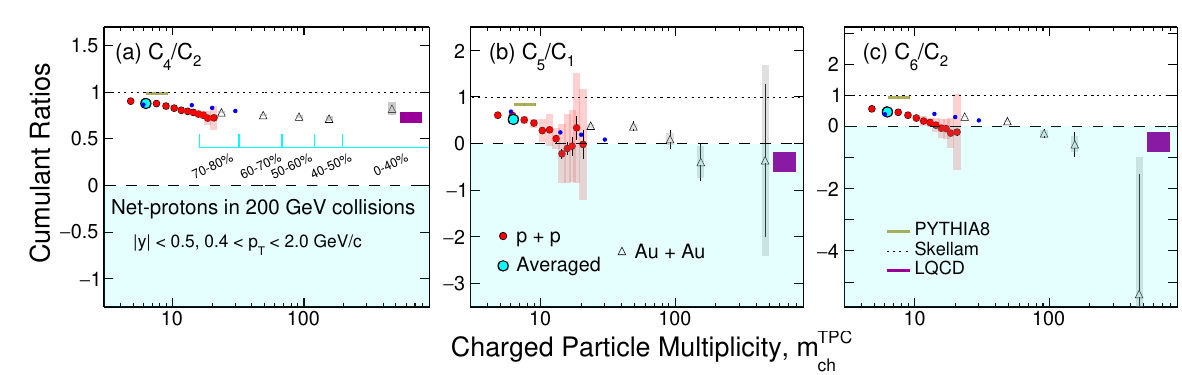}
\end{center}
\vspace{-0.5cm}
\caption{
Net-proton cumulant ratios as a function of reference multiplicity. Blue points represent the results from the numerical calculations incorporating Poisson fluctuations with the baryon number conservation.
}
\label{fig:Fig2_sup}
\end{figure}

\begin{enumerate}
    \item Generate baryons and antibaryons based on the Poisson distribution, where their mean values are taken from PYTHIA8 calculations for the reference multiplicity $m_{\rm ch}=6$. The $m_{\rm ch}$ is defined within the same acceptance for the experimental analysis.
    \item Events are proceeded only if the net-baryon number is 2, to incorporate the baryon number conservation.
    \item Produced (anti)baryons are randomly sampled based on the acceptance factor with the binomial response to determine the (anti)protons in the experimental acceptance. The acceptance factor is determined for $|y|<0.1$, $0.3$, and $0.5$, by  PYTHIA8 calculations.
    \item The above 1)-3) are repeated for many events, then we calculate net-proton $C_{4}/C_{2}$, $C_{5}/C_{1}$, and $C_{6}/C_{2}$.
    \item 1)-4) are repeated by varying the reference multiplicity for $m_{\rm ch}=14$, $20$, and $30$. Note that $m_{\rm ch}=20$ corresponds to the highest multiplicity that we can access in our experimental data.
\end{enumerate}
The results are shown in Fig.~\ref{fig:Fig1_sup}. Different lines/markers represent the reference multiplicity $m_{\rm ch}$. Each line has three points, corresponding to $|y|<0.1$, $0.3$, and $0.5$ from left to right. It is seen that the results from different $m_{\rm ch}$ are mostly on the same trend, except for $m_{\rm ch}=6$. This is because the mean values of (anti)protons within the rapidity acceptance is much smaller for $m_{\rm ch}=6$ compared to the higher $m_{\rm ch}$ events. All ratios decrease with increasing the acceptance factor. We find that, at the highest multiplicity events in the experiment ($m_{\rm ch}=20$), the ratios are $C_{5}/C_{1}\approx0.08$ and $C_{6}/C_{2}\approx0.2$.  These ratios are not negative even in the further higher multiplicity, $m_{\rm ch}=30$.
In Fig.~\ref{fig:Fig2_sup}, we overlaid these results by blue markers on Fig.~\ref{fig:plot1_ppnetp_March2021}. It is found that the slope of the decreasing trend seen in data (red) is much steeper for $C_{4}/C_{2}$ and $C_{6}/C_{2}$ than the effect of baryon number conservation (blue). We note that the detector efficiency present in $m_{\rm ch}^{\rm TPC}$ is 70-80\%, which does not still explain the slope difference.

\end{document}